# Enhancing high harmonic generation in a short-pulse two-color laser field by controlling the atomic-electron subcycle detachment and acceleration dynamics

O.V. Meshkov[1,2], M.Yu. Emelin[1], and M.Yu. Ryabikin[1,2,3]

[1]*Institute of Applied Physics, Russian Academy of Sciences, Nizhny Novgorod 603950, Russia*
[2]*Lobachevsky State University of Nizhny Novgorod, Nizhny Novgorod 603950, Russia*
[3]*Prokhorov General Physics Institute, Russian Academy of Sciences, Moscow 119991, Russia*

We present a study of the possibility to significantly enhance the efficiency of high-order harmonic generation (HHG) using few-cycle optical waveforms obtained by superposing two laser pulses of different color delayed optimally relative to each other. Special attention is paid to the dynamics of the depopulation of atomic states, which, one the one hand, promotes electrons to the continuum to take part in the high-energy photon emission, but, on the other hand, depletes the nonlinear medium. The use of the waveforms proposed here gives extra flexibility to control both the bound-state depopulation and the electron acceleration in the continuum. We demonstrate that the approach proposed here allows to increase by up to order of magnitude the efficiency of optical frequency conversion into sub-keV or few-keV photon energy ranges. High efficiency of HHG in optimal conditions is explained by the peculiarities of the photoelectron dynamics, which are in this case characterized by a combination of high-probability ejection of the electron responsible for the highest-order harmonic production and its subsequent strong acceleration accompanied by a relatively low probability of the bound-state depletion during the time interval between ionization and recollision.

*emelin@ufp.appl.sci-nnov.ru

**Keywords:** high-harmonic generation, few-cycle pulses, two-color fields, ionization.

## I. INTRODUCTION

With the advent of high-power midinfrared femtosecond lasers, new perspectives in the research of strong-field laser-matter interactions have been opened up. One of the most impressive benefits of using mid-IR sources to drive the processes associated with laser-induced ionization is a dramatic extension of the plateau in the spectrum of photons produced via high-order harmonic generation (HHG) in gases [1, 2]. When driven by intense mid-IR laser pulses, HHG has been shown to provide a tabletop source of coherent high-frequency light with keV photon energies [3] reaching, potentially, tens keV [4]. These enormous bandwidths, in the Fourier limit, are sufficient to support few-attosecond to subattosecond pulses.

HHG in gases is a three-step process, in which, according to the semiclassical model [5, 6], the electron is optical field-ionized, accelerated by the oscillating electric field, and driven back to the parent ion to emit a high-energy photon. The spectrum of HHG is plateau-like with the cutoff energy determined by the sum of the atomic ionization potential $I_\text{p}$ plus the electron's kinetic energy $3.17 U_\text{p}$ at the recollision instant [5, 7], where $U_\text{p} = e^2 E_0^2 / 4 m \omega_0^2$ is the average energy of electron oscillations in an ac electric field of amplitude $E_0$ and frequency $\omega_0$ (ponderomotive energy). The extension of the plateau produced with longer-wavelength driving lasers is due to the proportionality of the electron ponderomotive energy to the square of the laser wavelength. However, the efficiency of harmonic generation by individual atoms is known to scale unfavorably with the laser wavelength (typically, as $\lambda^{-\mu}$ with μ=5-6 [8-11]); this is caused to a large degree by a decrease in the probability of electron recollision with the parent ion due to the larger free-electron wave-packet spreading in a lower-frequency laser field.

Fortunately, this decrease in the efficiency of single-atom nonlinear response can be substantially compensated for by implementing phase-matched HHG at high gas pressures and large interaction lengths using gas-filled waveguides [3]. The capability of this approach is, however, limited by the requirement of negligible plasma dispersion, i.e., of very low ionization level of a gas, leading to limitations on the laser intensity to be used. A search for alternative approaches to increase harmonic yields is therefore a topical issue. One of the most promising approaches is to use multifrequency laser fields to facilitate favorable phase-matching conditions and/or to maximize harmonic emission from each particle of a gas. Recently, different schemes were proposed to achieve these goals using two- or three-color driving fields [12–16]. Most of these approaches relied on the use of many-cycle laser pulses; the physical origin of the yield enhancement and plateau widening on a single-atom level was usually explained in terms of the peculiarities of the free-electron behavior in a multicolor field.

In this paper, we study the possibility to enhance the efficiency of HHG using few-cycle optical waveforms obtained by superposing two laser pulses of different color delayed optimally relative to each other. A wide range of laser intensities is addressed. Special attention is paid to the subcycle dynamics of the depopulation of atomic states, which, on the one hand, promotes electrons to the continuum to take part in the high-energy photon emission, but, on the other hand, depletes the nonlinear medium [17, 18]. The use of waveforms addressed here gives extra flexibility, compared to the cases of single-color or many-cycle multicolor fields, in terms of the control over both the atomic bound-state depopulation and the electron acceleration in the continuum. To calculate the high-harmonic spectra, we use an analytical quantum-mechanical treatment of HHG within the strong-field approximation modified properly to adequately take into account the depletion of atomic states.

## II. THEORETICAL METHOD

In the strong-field approximation [19], the induced dipole moment of the atom in a strong low-frequency laser field linearly polarized along the $x$ axis can be represented as (hereinafter, atomic units are used)

$$x(t) = i\int_0^t d\tau \left(\frac{\pi}{\varepsilon + i\tau/2}\right)^{3/2} d_x^*\left(p_{st}(t,\tau) - \frac{A_x(t)}{c}\right) d_x\left(p_{st}(t,\tau) - \frac{A_x(t-\tau)}{c}\right),$$
$$\times E(t-\tau)\exp[-iS_{st}(t,\tau)]a^*(t)a(t-\tau) + c.c.,$$
(1)

where c.c. denotes complex conjugation; $E(t)$ and $\mathbf{A}(t)$ are the magnitude of the electric field and the vector potential of the laser pulse, respectively; $\varepsilon$ is the regularization parameter that can be chosen small; $\tau$ is the time of electron's free motion in the laser field after transition to the continuum;

$$p_{st}(t,\tau) = \frac{1}{\tau}\int_{t-\tau}^{t}\frac{A_x(t')}{c}dt'$$
(2)

is the stationary point, $\nabla_\mathbf{p} S(\mathbf{p},t,\tau) = 0$, of the quasiclassical action

$$S(\mathbf{p},t,\tau) = \int_{t-\tau}^{t}\left[\frac{1}{2}\left(\mathbf{p} - \frac{\mathbf{A}(t')}{c}\right)^2 + I_p\right]dt'$$
(3)

that describes the free motion of the electron in the laser field; the stationary value of the electron canonical momentum $\mathbf{p}$ given by Eq. (2) corresponds to the electron trajectory starting at the nucleus at $t-\tau$ and returning to the same position at $t$; $d_x(\mathbf{p})$ is the $x$ component of the dipole matrix element corresponding to the transition from the ground state to the continuum, which, for a hydrogenlike system with ionization potential $I_p$, takes the form

$$\mathbf{d}(\mathbf{p}) = i \frac{2^{7/2}(2I_p)^{5/4}}{\pi} \frac{\mathbf{p}}{(\mathbf{p}^2 + 2I_p)^3}. \qquad (4)$$

In Eq. (1), the integrands $d_x(p_{st}(t,\tau) - A_x(t-\tau))E(t-\tau)$ and $d_x^*(p_{st}(t,\tau) - A_x(t))$ determine, respectively, the probability amplitudes for the transition from the ground state to the continuum at $t-\tau$ and from the continuum back to the ground state at $t$. The consideration of the ground-state depletion is reducible to the multiplication of these factors by the amplitude $a(t-\tau)$ and $a^*(t)$ of the ground state at $t-\tau$ and $t$, respectively [19-22]. This amplitude can be approximated by

$$a(t) = \exp\left[-\int_0^t \frac{W(t')}{2} dt'\right], \qquad (5)$$

where $W(t)$ is the time-dependent ionization rate. Different approximations for $W(t)$ were used in Refs. [19-22]. In this article, we calculated $W(t)$ using the following analytical formula for the rate of tunneling ionization in a static field, adjusted for the barrier-suppression regime [23]:

$$W(t) = C_l^2 \left(\frac{4I_p}{|E(t)|}\right)^{\frac{2Z}{\sqrt{2I_p}}-1} \exp\left(-\frac{2(2I_p)^{3/2}}{3|E(t)|}\right) \exp\left(-\frac{6Z^2|E(t)|}{I_p(2I_p)^{3/2}}\right). \qquad (6)$$

In (5), $C_l$ is the constant determining the amplitude of the asymptotic wave function of the free electron at large distances from the nucleus; $Z$ is the charge of the atomic residue.

### III. RESULTS AND DISCUSSION

All the calculations presented below were carried out for a hydrogen atom exposed to a combination of two delayed ultrashort laser pulses of different colors. Both frequency components of the laser field are assumed to be linearly polarized along the $x$ axis. The total electric field of the synthesized waveform is given by

$$E(t) = E_1 \cos[\omega_1(t+\tau)+\phi]\exp\left[-2\ln 2\frac{(t+\tau)^2}{\tau_1^2}\right] + E_2 \cos(\omega_2 t)\exp\left(-2\ln 2\frac{t^2}{\tau_2^2}\right), \qquad (7)$$

where $E_1$, $\omega_1$, $\tau_1$ and $E_2$, $\omega_2$, $\tau_2$ are the amplitude, frequency, and duration of the 1st and 2nd pulse, respectively; $\tau$ is the delay between two pulses and $\phi$ is an additional phase shift of the 1st pulse. The parameters of the waveform (7) were optimized to ensure the maximum harmonic yield within a given range of harmonic photon energies. The harmonic yield to be maximized is proportional to the integral

$$P = \int_{\omega_{\min}}^{\omega_{\max}} |\ddot{x}_\omega|^2 d\omega, \qquad (8)$$

where $\ddot{x}_\omega$ is the Fourier component of the electron dipole acceleration $\ddot{x}(t)$. The optimization of the parameters of the waveform (7) was performed both by iterating over parameter values within their specified intervals and using genetic algorithms.

One of the main differences between our research and most of other studies aimed at finding the optimal conditions for HHG driven by multicolor fields is that we allowed the atomic bound-state population to be varied in accordance with Eqs. (5)-(6). The main parameter that controls this population is the peak intensity of the laser pulse. As for the limits on this parameter, notice that in the description on a microscopic (single-atom) level taking into account the dynamics of atomic-state populations, the limitations on the laser intensity, as a rule, automatically follow from the fact that at sufficiently high intensities, there is significant depletion of atomic bound states, which reduces the efficiency of frequency conversion by atomic emitters via the recombination mechanism [17, 18]. In practice, the laser intensities usable for HHG are even lower and are limited by macroscopic effects leading to the requirement that the concentration of the emerging plasma or, in other words, the ionization level of the medium does not exceed some critical level [24]. Notice, however, that, owing to the sharp increase of the ionization rate with laser intensity, the critical intensities determined by the above two criteria are not that different. Moreover, in recent theoretical and experimental studies [25, 26], the "overdriven" regime of HHG has been demonstrated, in which a high-flux harmonic generation in the water window (284 to 540 eV) was obtained using few-cycle laser pulses with higher peak intensities than are typically chosen to avoid strong ionization of the medium. The mechanism explaining this observation is lensing caused by emerging plasma, which changes the laser beam shape rapidly, leading to strong spatial and temporal gradients in intensity and phase and, ultimately, to transient subcycle phase matching. In these conditions, the maximum values of allowable laser intensities may approach those dictated by a single-atom response.

Since in our case the macroscopic description is extremely computation demanding, especially given the multi-parametric nature of the problem, in this study we restrict ourselves to a microscopic treatment, aiming to optimize the performance of an elementary atomic emitter. However, in this work we separately consider two regimes related to the cases of (i) high and (ii) moderate intensities. In the first case, we do not impose additional restrictions on the laser intensities, allowing them to be as limited by the depletion of the nonlinear medium. In the second case, we introduce an additional restriction, which consists in the requirement that the level of ionization of the medium resulting from its interaction with the laser field does not exceed a few percent, which is an ionization level usually taken as acceptable in terms of phase matching conditions.

From the calculations for different energy ranges of the harmonic emission, we found that, typically, the harmonic yield is maximized provided that the peak intensities of two laser pulses with different carrier frequencies differ significantly in magnitude. More specifically, the longer-wavelength component ($\lambda_2$) in the optimal case has, as a rule, a significantly higher intensity than its shorter-wavelength ($\lambda_1$) companion. In view of this, the longer- and shorter-wavelength pulses can be called, respectively, the main and control ones.

TABLE I. Summary of optimal values of laser parameters maximizing the harmonic yield in different spectral intervals in the high-intensity regime. Right column shows the enhancement factor $W$ due to the use of a two-color driving field instead of a single-color one.

| Interval (eV) | $I_1$ ($10^{14}$ W/cm$^2$) | $I_2$ ($10^{14}$ W/cm$^2$) | $\lambda_1$ (μm) | $\lambda_2$ (μm) | $\tau$ ($T_2$) | $\phi$ (rad) | $W$ |
|---|---|---|---|---|---|---|---|
| 283-543 | 0,75 | 4,75 | 0,8 | 1,8 | 0,95 | 0 | 2,8 |
| 543-750 | 0,75 | 4,5 | 1 | 2,4 | 0,95 | 0 | 4,2 |
| 750-1000 | 1,25 | 3,75 | 0,8 | 3,2 | 0,95 | 0 | 5,5 |
| 1000-1500 | 0,5 | 4,75 | 1,4 | 3,2 | 0,95 | 0 | 5 |
| 1350-1550 | 0,5 | 6,5 | 1,6 | 3,2 | 0,75 | π | 5,5 |

Table 1 presents the results of calculations for the high-intensity regime. Shown are the optimal values of the parameters maximizing the harmonic yield in different spectral intervals. In the optimization procedure, the peak intensity of the pulses $I_1$ and $I_2$ ranged from $5\times10^{13}$ W/cm$^2$ to $1.55\times10^{15}$ W/cm$^2$ with a step of $2.5\times10^{13}$ W/cm$^2$; wavelengths $\lambda_1$ and $\lambda_2$ ranged from 0.8 μm to 3.2 μm with a step of 0.2 μm. The delay $\tau$ varied from zero to one cycle ($T_2$) of a longer-wavelength component, with a step of 0.05 cycle. Pulse durations $\tau_1$ and $\tau_2$ were fixed and amounted to 1.5 cycles of the corresponding field. Phase $\phi$ could take values from 0 to π. The quantity denoted as $W$ in the last column of Table 1 is the enhancement factor defined as the ratio of the harmonic yields in the cases of two-color and one-color laser fields. Note that, as it turns out, for each spectral interval considered, there is also another possible choice of parameters, besides the one presented in Table 1, which provides approximately the same enhancement factor; the optimal delay $\tau$ in this case is 0.25 $T_2$ for the energy interval of 1350-1550 eV and 0.55 $T_2$ for the remaining intervals, and the optimal values of the other parameters are slightly different from those given in Table 1. In both cases, as the integration window moves towards higher photon energies, the enhancement factor tends to increase, whereas the optimum wavelength ratio $\lambda_1:\lambda_2$ tends to 1:2.

Next, we consider in more detail the case of the highest energy interval of those presented in Table 1 in order to look for conditions to achieve an even higher harmonic yield. We fix the wavelengths of the driver frequency components equal to $\lambda_1 = 1.6$ μm and $\lambda_2 = 3.2$ μm, and for delay $\tau$ fixed at 0.25 $T_2$ or 0.75 $T_2$ we vary the duration $\tau_1$ of the control pulse (for fixed duration $\tau_2$ of the main pulse), as well as the intensity of both pulses. As a result, we found that with optimal values of parameters equal to $I_1=8\times10^{13}$ W/cm$^2$, $I_2=6.4\times10^{14}$ W/cm$^2$, and $\tau_1=2.4\,T_1$, the harmonic yield in the interval under consideration is 6.9 times higher than when using only the main pulse.

Figure 1 shows in detail the results obtained with the optimal parameters found above. Shown are the time dependences of the laser electric field (Fig. 1(a)) and the atomic ground-state population (Fig. 1(c)), as well as the harmonic emission spectra (Fig. 1(d)), for a single pulse (magenta line) and the synthesized two-pulse waveform (black line). Also shown are the time profiles of the frequency constituents of a two-color waveform (Fig. 1(b)).

To explain the origin of the high-harmonic yield enhancement under the conditions indicated above, we first note that the main events leading to the generation of high harmonics, whose spectrum is presented in Fig. 1(d), occur in the time interval marked with a dashed line in Figs. 1(a-b); this can be seen from the inspection of the dynamics of atomic ground-state population shown in Fig. 1(c). In this time interval, the electric field has two extrema of opposite signs. Around the first extremum, there is a high probability of electron ejection. Once ejected, according to the three-step model, the electron is thrown by the field away from the atomic core, and further, after the field reverses its direction, it is accelerated back to the parent ion to encounter it near the time when the field goes to zero, causing, with some probability, a harmonic photon emission. In addition, during the second half oscillation of the field, a further depopulation of atomic levels may occur, which, in the case of large field strength at the corresponding extremum, can result in greatly reduced high-frequency dipole moment due to vanishing ground-state amplitude $a^*(t)$, see Eq. (1).

From the time profiles of the electric field shown in Figs. 1(a-b), it can be seen that, in the two-color case, the optimal choice of central wavelengths of two pulses and a delay between them corresponds to such a characteristic relationship between the phases of the field oscillations of the waveform constituents, for which the oscillations of two fields in the time interval discussed here are added in phase near the first extremum, whereas near the second extremum they are in antiphase.

The change in the waveform of the electric field introduced by adding a control field results in a significant change in the dynamics of the active electron. The latter refers both to the bound-state depopulation and to the population transfer to the manifold of continuum states. For further discussion, it is convenient to present the electron wave function simplistically as a superposition of the bound ($\psi_b$) and free ($\psi_f$) parts: $\psi = c_b\psi_b + c_f\psi_f$. Then the average value of the dipole moment $\boldsymbol{\mu}(t) = \langle\psi(\mathbf{r},t)|-\mathbf{r}|\psi(\mathbf{r},t)\rangle$ we are interested in can be written as

$$\boldsymbol{\mu} = |c_b|^2\langle\psi_b|-\mathbf{r}|\psi_b\rangle + c_b^*c_f\langle\psi_b|-\mathbf{r}|\psi_f\rangle + c_bc_f^*\langle\psi_f|-\mathbf{r}|\psi_b\rangle + |c_f|^2\langle\psi_f|-\mathbf{r}|\psi_f\rangle. \quad (9)$$

The high-frequency part of the induced dipole moment is defined predominantly by the second and third terms on the right-hand side of (9) – see, for example, [27]. From (9), it is easy to see that for the existence of a strong high-frequency single-atom response, a sufficiently large population of both bound and free states (a large absolute value of product $c_b^*c_f$) is necessary.

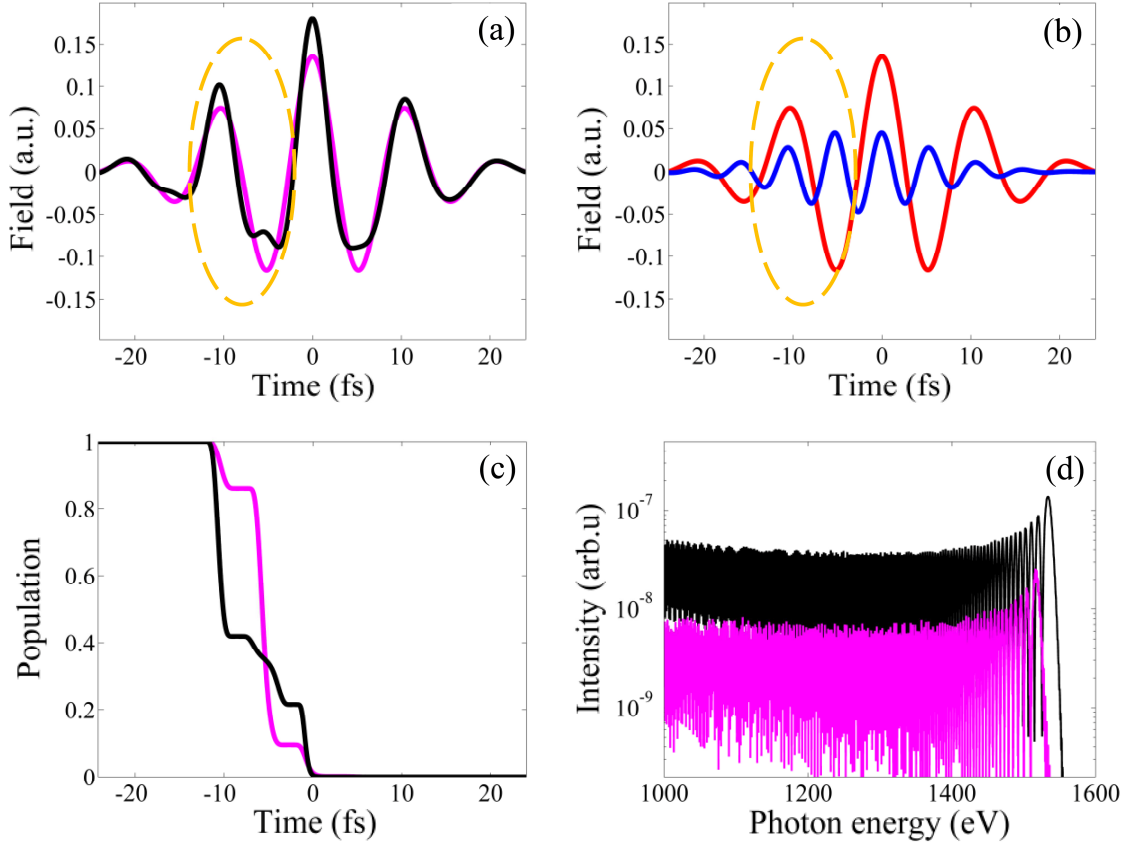

FIG. 1. (a) Time profile of the laser electric field, (b) components of a two-color field, (c) time dependence of the atomic ground-state population, and (d) high-energy part of the HHG spectrum for a hydrogen atom driven by a single-color ($\lambda$=3.2 μm, magenta line) and two-color ($\lambda_2$=3.2 μm, $\lambda_1$=1.6 μm, black line) few-cycle laser pulse. FWHM of the 3.2 μm pulse is 1.5 cycles; the values of other parameters (see text) are chosen to maximize the harmonic yield between 1350 and 1550 eV: $I$=6.4×10$^{14}$ W/cm$^2$ (single-color field), $I_2$=6.4×10$^{14}$ W/cm$^2$, $I_1$=8.0×10$^{13}$ W/cm$^2$, $\tau_1$=2.4 $T_1$, $\tau$=0.25 $T_2$, $\phi$=π (two-color field); $T_{1,2}$=$\lambda_{1,2}/c$. Dashed line outlines the region responsible for the formation of the spectra shown.

As follows from Fig. 1(c), both in the single-color and two-color cases, a rather strong atomic ionization occurs near the time t ≈ -10 fs, which corresponds to the local maxima of the electric field in both cases. Note, however, that the probability of ionization near this time in the two-color case is significantly higher; therefore, in this case, the probability of finding an electron in the continuum states is higher (the $c_f$ value is larger). Further electron dynamics in these two cases are also different. In the single-color case, the field in the next half-cycle after the first ionization event is even stronger than in the previous one. Hence, by the time t ≈ - 2 fs, when the free part of the electron wave packet revisits the parent ion, the ground state is strongly depleted by the field (the $c_b$ value is small) (see Fig. 1(c), magenta curve), which leads to the suppression of the high-harmonic production. On the contrary, in the two-color case, the field in the half-cycle after the first ionization event is weaker in amplitude than in the previous half-oscillation, hence, it produces less additional ionization (see Fig. 1(c), black curve). On the other hand, this half-oscillation turns out to last longer, and by the time of the return, the free electron has time to gain similar or even higher energy than in the single-color case. Since in the time interval discussed here, the population of both free and bound states in the two-color case is large enough, the high-frequency polarization is stronger and the resulting high-harmonic signal is much more intense than in the single-color case (see Fig. 1(d), black curve). Summarizing, we can say that in the case of optimized two-color waveform, both the electron release from the atom and its acceleration in free space follow a more favorable scenario in terms of the induced polarization.

Bearing in mind the effectiveness of the proposed two-color scheme shown above by the example of enhanced production of high-order harmonics in the photon energy range from 0.3 to 1.5 keV, we further study the possibility of extending this scheme to even higher photon energies using shorter-wavelength drivers available today. To this end, we carried out calculations similar to those performed above, now for the case of an ultrashort waveform synthesized from few-cycle pulses with a central wavelength of 3.9 μm [28] and its second harmonic. The results are shown in Table 2. The data presented clearly indicate that the gain from using two-color drivers to generate high-energy photons via HHG increases with the energy of the photons to be produced.

TABLE II. Summary of optimal values of laser parameters ($I$, single-color field; $I_1$, $I_2$, and $\tau_1$, two-color field) maximizing the harmonic yield in different spectral intervals in the high-intensity regime for single-color ($\lambda$ = 3.9 μm) and two-color ($\lambda_1$ = 1.95 μm and $\lambda_2$ = 3.9 μm) cases. In both cases, the pulse duration of the long-wavelength component is fixed at 1.5 cycles. For the two-color field, the delay $\tau$ and phase $\phi$ are fixed at $\tau$=0.25 $T_2$ and $\phi$=π, respectively. Right column shows the enhancement factor $W$ due to the use of a two-color driving field.

| Interval (eV) | $I$ ($10^{14}$ W/cm$^2$) | $I_1$ ($10^{14}$ W/cm$^2$) | $I_2$ ($10^{14}$ W/cm$^2$) | $\tau_1$ ($T_1$) | $W$ |
|---|---|---|---|---|---|
| 1700-1900 | 4 | 0,75 | 5,25 | 2,7 | 4,8 |
| 1900-2100 | 4,4 | 0,5 | 6 | 2,7 | 7,8 |
| 2100-2300 | 6,5 | 0,75 | 6,5 | 2,1 | 8,1 |
| 2300-2500 | 7,1 | 2 | 7 | 1,8 | 8,6 |
| 2500-2700 | 7,6 | 1,25 | 7,5 | 1,5 | 9,1 |
| 2700-2900 | 8,2 | 1,25 | 8 | 1,5 | 10 |
| 2900-3100 | 8,7 | 1,5 | 8,5 | 1,5 | 11,7 |

It should be noted, however, that with an increase in the energy of the generated photons, this gain is achieved at an increasingly higher price, since more intense and short laser pulses are required to obtain this gain, as can be seen from the data shown here. On the other hand, more detailed study shows that the duration of the control pulse is of no critical importance. In particular, in the limit of infinitely long control pulse, the enhancement factor $W$ for photon production with energy up to 2.1 keV turns out to be within 0.85 of its value for optimally chosen pulse duration.

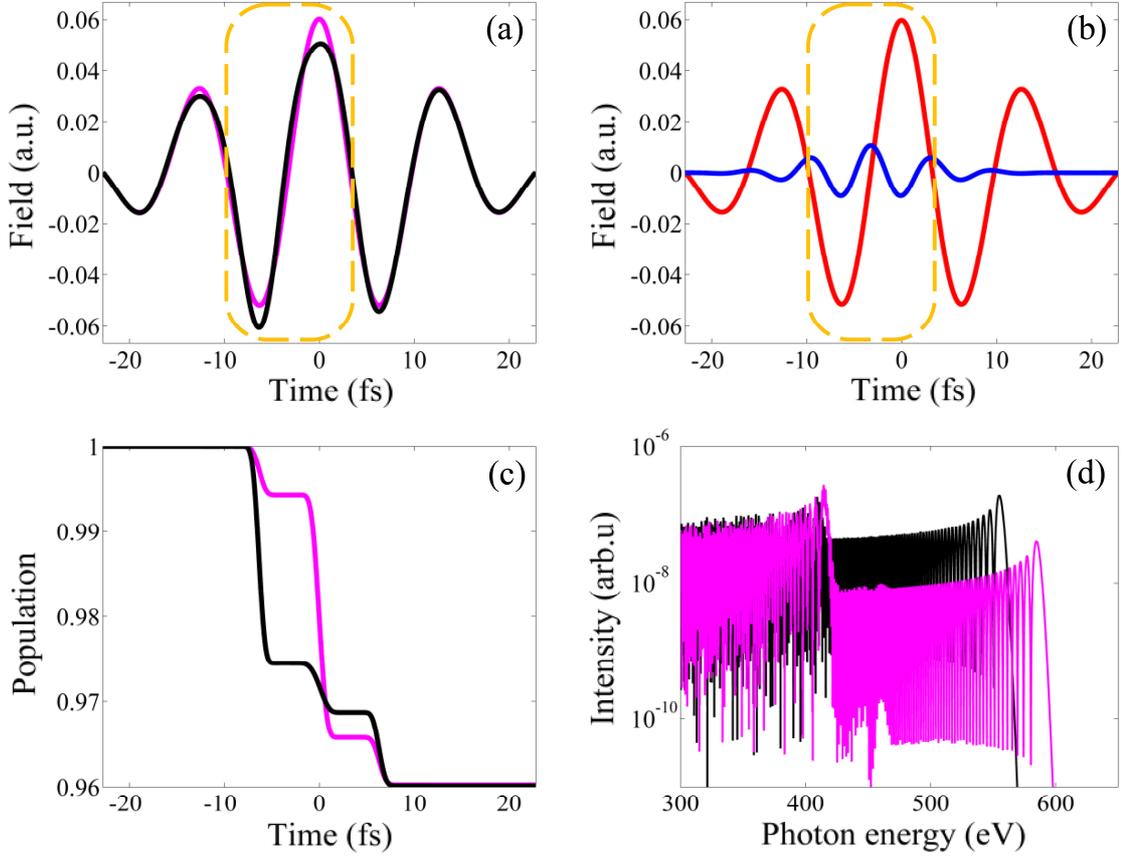

FIG. 2. (a) Same as in Fig. 1 but for a single-color ($\lambda$=3.9 μm, pink line) and two-color ($\lambda_2$=3.9 μm, $\lambda_1$=1.95 μm, black line) few-cycle laser pulse. The ionization level is restricted to 4%. FWHM of the 3.9 μm pulse is 1.5 cycles; the values of other parameters are chosen to maximize the harmonic yield between 500 and 600 eV: $I$=1.27×10$^{14}$ W/cm$^2$ (single-color field), $I_2$=1.25×10$^{14}$ W/cm$^2$, $I_1$=4.0×10$^{12}$ W/cm$^2$, $\tau_1$=1.5 $T_1$, $\tau$=0.25 $T_2$, $\phi$=0 (two-color field).

As mentioned above, we also considered the case of moderate-intensity laser fields. In this case, we restricted ourselves to considering the case when the level of laser-induced ionization of the medium does not exceed 4 %, which is usually acceptable for maintaining phase matching. We found that even with this limitation, it is still possible to optimize the electron dynamics in such a way as to obtain a harmonic yield several times higher than in the case of a single-color field. A typical example is shown in Fig. 2, where the results are shown for the set of parameters $I_2$=1.25×10$^{14}$ W/cm$^2$, $I_1$=4.0×10$^{12}$ W/cm$^2$, $\tau_1$=1.5 $T_1$, $\tau$=0.25 $T_2$, and $\phi$=0. With this choice of the parameters of a two-color pulse, the harmonic yield between 500 and 600 eV is maximal. For comparison, the case of a single-color pulse is also presented, in which, thanks to the similar peak intensity of the field, the ultimate level of ionization of the medium is the same. In the optimal two-color case, the subcycle field evolution (see Fig. 2(a), black curve) turns out to be modified so that the bound-state depopulation follows a more favorable scenario (Fig. 2(c), black curve) for obtaining an intense high-frequency atomic polarization response. More specifically, in the two-color case, the production of the electrons responsible for the generation of near-cutoff harmonics is much more efficient, whereas the subsequent nonlinear medium depletion is much less pronounced than in the single-color case. As a result, the integral yield of the harmonics in the energy window 500-600 eV in the two-color case is 3.8 times higher, whereas the spectral intensities in the 420-555 eV range are 6-10 times higher compared to the single-color case.

## IV. CONCLUSION

In conclusion, our study shows that not only the trajectories of electrons in the continuum but also the subcycle behavior of the population transfer from the atomic bound states into the continuum can be finely controlled in order to induce more efficiently the high-frequency nonlinear polarization response of the gaseous media driven by the femtosecond laser pulses. This control can be achieved using the ultrashort optical waveforms tailored from a pair of few-cycle pulses of different color properly delayed relative to each other. An important ingredient of the proposed scheme is the ability to promote a sufficient share of electrons to the continuum followed by their efficient acceleration without further significant depletion of the nonlinear medium. In the optimal case, the latter is achieved if stronger acceleration of the electron is ensured by elongation of the accelerating half-oscillation of the field rather than by increasing its amplitude. The calculation results demonstrate usability of the proposed scheme to increase by up to order of magnitude the efficiency of optical frequency conversion into sub-keV and few-keV photon energy ranges compared to a single-color case.

## ACKNOWLEDGMENTS

We acknowledge financial support from the Ministry of Science and Higher Education of the Russian Federation (state assignment for the Institute of Applied Physics RAS, project No. 0035-2019-0012). The study of atomic ionization dynamics was supported by Russian Foundation for Basic Research (Grant No. 18-02-00924).

___________________________________________________________________